\documentclass[superscriptaddress,amsmath,amssymb,prx,nofootinbib,twocolumn]{revtex4-2}
\usepackage{hyperref}
\usepackage[graphicx]{realboxes}
\usepackage{placeins} 
\usepackage{babel}
\usepackage{times}
\usepackage[usenames,dvipsnames]{color}
\usepackage{soul}
\usepackage[utf8]{inputenc}
\usepackage{float}
\usepackage{dsfont} 
\usepackage{mathtools} 

\begin{document}

\newcommand{\REF}[1]{\textcolor{red}{REF(#1)}}
\newcommand{\red}[1]{\textcolor{red}{#1}}
\newcommand{\TODO}[1]{\textcolor{red}{TODO #1}}

\newcommand{\bra}[1]{\left\langle #1 \right\vert }
\newcommand{\ket}[1]{\left\vert #1 \right\rangle }
\newcommand{\ev}[1]{\left\langle #1 \right\rangle }

\newcommand{\sx}{\sigma^x}
\newcommand{\sz}{\sigma^z}

\newcommand{\vecr}{\mathbf{r}}
\newcommand{\veck}{\mathbf{k}}
\newcommand{\pipi}{\left(\frac{\pi}{2}, \frac{\pi}{2}\right)}

\newcommand{\Cth}{C_3^h}
\newcommand{\Ctk}{C_3^k}
\newcommand{\Cf}{C_4}

\newcommand{\id}{\mathds{1}}

\title{Scattering and induced false vacuum decay in the two-dimensional quantum Ising model}

\author{Luka Pave\v{s}i\'{c}}
\email{luka.pavesic@unipd.it}
\affiliation{Dipartimento di Fisica e Astronomia “G. Galilei”, Universit\`a degli Studi di Padova, via Marzolo 8, I-35131 Padova, Italy}
\affiliation{Istituto Nazionale di Fisica Nucleare (INFN), Sezione di Padova, I-35131 Padova, Italy}

\author{Marco Di Liberto}
\affiliation{Dipartimento di Fisica e Astronomia “G. Galilei”, Universit\`a degli Studi di Padova, via Marzolo 8, I-35131 Padova, Italy}
\affiliation{Istituto Nazionale di Fisica Nucleare (INFN), Sezione di Padova, I-35131 Padova, Italy}
\affiliation{Padua Quantum Technologies Research Center, Universit\`a degli Studi di Padova, Padova, Italy}

\author{Simone Montangero}
\affiliation{Dipartimento di Fisica e Astronomia “G. Galilei”, Universit\`a degli Studi di Padova, via Marzolo 8, I-35131 Padova, Italy}
\affiliation{Istituto Nazionale di Fisica Nucleare (INFN), Sezione di Padova, I-35131 Padova, Italy}
\affiliation{Padua Quantum Technologies Research Center, Universit\`a degli Studi di Padova, Padova, Italy}

\begin{abstract}
We study scattering in the quantum Ising model in two dimensions.
In the ordered phase, the spectrum contains a ladder of bound states and intertwined scattering resonances, which enable various scattering channels.
By preparing wave packets on a $24 \times 24$ lattice and evolving the state with tensor networks, we explore and characterize these regimes, ranging from elastic scattering in the perturbative regime, to non-perturbative processes closer to the critical point.
Then, we break the spin inversion symmetry and study the stability of the metastable false vacuum state on the collision of its excitations. We find that a highly-energetic scattering process can induce a violent decay of the false vacuum, and investigate the spread of the resulting true vacuum bubble. 
\end{abstract}

\maketitle

Scattering experiments are a paradigmatic tool for probing the inner structure of Nature, from high-energy to condensed matter physics.
Yet, reconstructing the underlying theory from the products of scattering experiments is not always possible, especially in strongly interacting theories where perturbative methods fail.
Real-time numerical simulations of scattering thus play an important role by providing direct access to the evolution of the interactions throughout the process.
However, classically simulating the dynamics of interacting quantum systems is often infeasible.  
Quantum simulators could overcome these limitations, and quantum simulations of scattering events are a realistic prospect in the near-term future.
See Refs.~\cite{Byrnes2006, Jordan2011, Banuls2020, Bauer2023, Su2024, DiMeglio2024} for roadmap papers, and Refs.~\cite{Surace2021, Belyansky2024, Ingoldby2025, Bennewitz2025, Chai2025, Schuhmacher2025, Davoudi2025, Farrell2025, Chai2025_fermionic, Papaefstathiou2025} for proposals and implementations of pioneering experiments.

Spin chains are the go-to models for scattering simulations, as they can be efficiently simulated with tensor networks~\cite{Vanderstraeten2014, VanDamme2021_scattering, Karpov2022}. 
More importantly, they also emerge as effective descriptions of more complex models and phenomena.
A prominent example is confinement, a non-perturbative mechanism that binds quarks into hadrons~\cite{Wilson1974}. 
Analogous phenomena exist in the Ising model in one~\cite{Kormos2016, Liu2019} and two dimensions~\cite{James2019, Balducci2022, Balducci2023, Tindall2024, pavesic2025}, where excitations feel a confining potential that binds them into bound states, reminiscent of mesons. 
Simulating scattering in spin chains can deepen our understanding of confining theories~\cite{Surace2021, Bennewitz2025}, and quantum field theory~\cite{Jordan2011, Milsted2022, Jha2024}. Understanding the nature of composite excitations is also important in strongly-correlated condensed matter, like high-temperature superconductors~\cite{Grusdt2018, Greiner2024}. 

Notably, real-time scattering simulations -- quantum or classical -- remain restricted to models in one spatial dimension (1D).
Compared to their higher-dimensional counterparts, these can be limited in the physics content they offer. Confinement, chirality, topological order and excitations, the magnetic phase of QED, all require dimensions higher than one~\cite{Fradkin2013}.
Furthermore, the constrained 1D geometry means that regimes with resonances for inelastic scattering might not be easily available~\cite{Surace2021, Rigobello2021, Bennewitz2025, Jha2024}.
Extending the simulations of scattering to higher spatial dimensions is thus critical for capturing the features of realistic theories in 3D.

Here, we study the scattering in the two-dimensional quantum Ising model. 
We employ tree tensor networks (TTN)~\cite{Tagliacozzo2009, Murg2010} to simulate the collisions of wave packets of elementary spin excitations -- \textit{magnons} -- on a two-dimensional lattice of $24 \times 24$ spins.
We uncover a rich interplay of interwoven scattering resonances and the resulting production and decay of composite particles and bound states.
Next, we tune the system into a symmetry-broken regime with stable and metastable vacua, and investigate the scattering in a metastable false vacuum. 
We uncover a dynamical regime where the scattering of two particles induces a violent decay of the false vacuum, a non-perturbative phenomenon that is expected to be exponentially suppressed in the semi-classical regime~\cite{Kuznetsov1997, Levkov2005, Demidov2015}.

Our results demonstrate that tensor networks can capture scattering dynamics and probe non-perturbative physics in 2D.
This opens the door to such simulations in more complex settings, like lattice-gauge theories and condensed matter systems, and paves the way for future quantum simulations in regimes beyond the reach of classical simulations.

\section{The model and its low-energy spectrum}
\label{sec:model}

\begin{figure*}
    \centering
    \includegraphics[width=\textwidth]{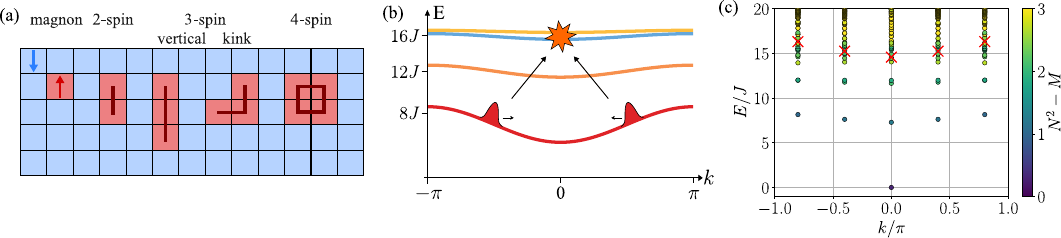}
    \caption{
    The spectrum of the 2D Ising model. 
    (a) Low-energy excitations above the polarized state are domains of flipped spins. Here we show the excitations up to three, and the lowest-energy four-spin excitation. 
    (b) A sketch of the single-excitation spectrum. The bands correspond to excitations shown in panel (a). We prepare wave packets in the magnon band (red). The scattering energy is resonant with the 3-spin and 4-spin bands. 
    (c) The many-body spectrum of a $5 \times 5$ lattice for $g/J = 1$ for momentum $\mathbf k = (k,k)$. The color indicates the change of magnetization with respect to the fully polarized state. Red crosses denote twice the energy of the one-spin excitation; this corresponds to the energy available in scattering.
    }
    \label{fig:excitations_and_spectrum}
\end{figure*}

The quantum Ising model on a 2D square lattice of $N \times N$ spins with periodic boundary conditions is described by the Hamiltonian
\begin{equation}
    H = -J \sum_{\langle ij \rangle} Z_i Z_j - g \sum_i X_i - h \sum_i Z_i,
\end{equation}
with positive interaction $J$, transverse field $g$ and longitudinal field $h$. $Z_i$ and $X_i$ are Pauli operators. The sum in the first term runs across nearest neighbors.

We focus on the ordered ferromagnetic phase, \mbox{$g < g_c \sim 3.04 J$}~\cite{Rieger1999}, with a two-fold degenerate ground state with spins polarized either in the $+Z$ or $-Z$ direction. 
The two subspaces are decoupled by the spin-inversion symmetry, which is broken by the longitudinal field ($h \neq 0$). 

The elementary excitations above the polarized ground state are magnons; spin flips generated by the $X$ operator. 
Their excitation energy is $8 J$ for $g = h = 0$, and they acquire a finite dispersion at finite $g/J$.
At leading order the dispersion relation reads~\cite{Dusuel2010}:
\begin{equation}
    \label{eq:dispersion}
    \varepsilon_1 = 8J - \frac{g^2}{J} \left( \frac{1}{2} +\frac{1}{4} \left  ( \cos k_x+\cos k_y \right) \right) + \mathcal{O}(g^4)\ .
\end{equation}
In general, all even orders of $g$ contribute. 

Because it is energetically cheaper to flip spins next to an existing excited spin, higher-energy excitations form small domains of flipped spins~\cite{Sachdev2011}, acting as composite particles.
In Fig.~\ref{fig:excitations_and_spectrum}(a) we show excitations up to the four-spin domain, and the single-excitation spectrum is sketched in Fig.~\ref{fig:excitations_and_spectrum}(b). 
Hopping of a domain of $n$ spins is a process in $2n$-th order in $g$, so the bandwidth scales with $g^{2n}$. Larger excitations are thus heavier and less mobile.
The domains can have distinct shape, giving rise to multiple bands.
For example, the 2-spin excitations can be oriented either horizontally or vertically, and the two species are coupled in second order in $g$. 
As a consequence, their spectrum consists of two split and intertwined bands. 
The onset of this effect can be seen in Fig.~\ref{fig:excitations_and_spectrum}(c) where we show the many-body spectrum of a $5 \times 5$ lattice. 
Three-spin excitations display a similar, but much richer behavior.
There are two species of the 3-spin particle, a vertically/horizontally extended one, and a kink.  
They mix among each other in second order in $g$.
They also have the same perimeter as the 4-spin square excitation, which is thus resonantly coupled to the kinks at first order in $g$~\cite{Balducci2022, Balducci2023}. 

The red crosses in Fig.~\ref{fig:excitations_and_spectrum}(c) correspond to the energy of two magnons with equal but opposite momenta $\veck = \pm (k,k)$, showing that the two-magnon continuum is resonant with the three- and four-spin excitations.
Resonant transitions between these states drive the scattering processes we discuss next.

\section{From elastic to inelastic scattering}
\label{sec:scattering}

\begin{figure*}
    \centering
    \includegraphics[width=\textwidth]{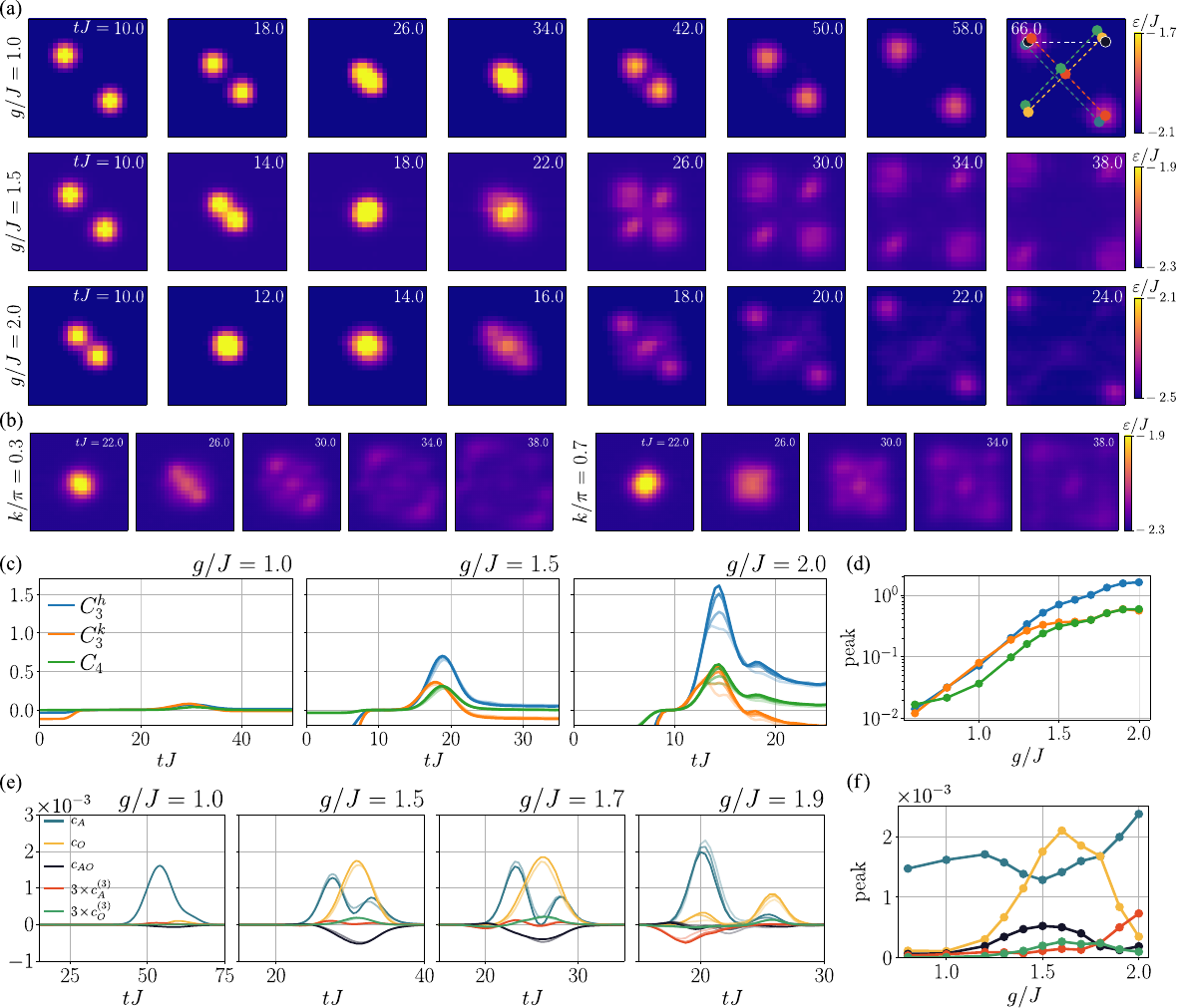}
    \caption{
    Scattering of magnon wave packets.
    (a) The evolution of energy density for varying $g/J$. The wave packets have momenta $\veck = \pm\pipi$.
    (b) The evolution of energy density at $g/J=1.5$ for two values of $\veck = \pm(k,k)$.   
    (c) The evolution of the quasiparticle correlators for representative values of $g/J$. The value at $tJ=10$ is subtracted to disentangle the scattering from the background effects. 
    (d) The $g$-dependence of the peak value of the correlators. 
    (e) The long-range correlations. The points where the correlations are measured are sketched in the top right panel of (a) using the same colors scheme. As they are smaller than the two-point measurements, the three-point correlators are multiplied by a factor of 3 for visibility. 
    (f) The $g$-dependence of the peak value of the long-range correlations.
    The curves drawn with different shades of the same color correspond to simulations with different bond dimension, the darkest being the largest. For $g/J = 1.0$ we use $\chi=100$ and $150$, for $g/J = 1.5$ $\chi = 100, 150, 200$, and for $g/J > 1.5$ we have $\chi = 100, 150, 200, 250$.
    }
    \label{fig:scattering}
\end{figure*}

We prepare a state with two wave packets at $g=0$ as a superposition of product states. Then, we adiabatically ramp up $g$ to the target value on the timescale $\tau = 10/J$, to generate accurate dressed states~\cite{Bennewitz2025}. See Sec.~\ref{sec:methods} for details. 
We simulate the scattering process at varying values of $g$. This modifies the energy, but more importantly, tunes the nature of the interaction; at small $g/J$ we expect perturbative interactions and approximately conserved total magnetization~\cite{Balducci2022, Balducci2023}, while the non-perturbative physics emerges at larger $g/J$.
The results are collected in Fig.~\ref{fig:scattering}.

To visualize the wave packets, we use the energy density
\begin{equation}
    \varepsilon_{i} \coloneqq -\frac{1}{2}J \sum_{j \mathrm{\ n.n.\ of\ } i} \langle Z_i Z_j \rangle - g \langle X_i \rangle - h \langle Z_i \rangle
\end{equation} 
with $j$ nearest neighbors of $i$.
Its evolution through the scattering process is shown in Fig.~\ref{fig:scattering}(a) for three representative values of $g/J$ and $\veck = \pm \pipi$.
The scattering process is predominantly elastic for $g/J = 1$ (top). At $g/J = 1.5$ (middle) it goes through an intermediate resonance, and we find signatures of production of a heavier particle for $g/J = 2.0$ (bottom).

One feature that stands out is the spatial distribution of the products: they dominantly disperse in two directions.
This is a consequence of the shape of the magnon dispersion.
The probability amplitude for scattering into a final state is proportional to the density of its states at the given energy, given by $\vert\nabla_\veck E\vert^{-1}$ for dispersion $E$.
The magnon dispersion, Eq.~\eqref{eq:dispersion}, produces a density of states with divergences -- known as van Hove singularities -- at $k_x = k_y = \pm\frac{\pi}{2}$ and $k_x = -k_y = \pm\frac{\pi}{2}$. This greatly increases the probability for scattering in these directions.
Note that this only applies if the scattering generates a pair of magnons.
This indicates that the process at $g/J \sim 1.5$ must be two magnons $\rightarrow$ intermediate excitation $\rightarrow$ two magnons.
The effect fades away at larger $g/J$, hinting at a different inelastic process in that regime. 

To quantify the particles produced during the scattering process, we measure correlators corresponding to approximate creation operators of composite excitations, built from products of the number operator $n_{i,j} = \frac{1}{2} \left(Z_{i,j} + \id_{i,j} \right)$ on neighboring lattice sites. 
We assume that the scattering products are constrained to the resonant subspace of particles with 8 domain walls, and measure the three rightmost excitations shown in Fig.~\ref{fig:excitations_and_spectrum}.

The three-spin horizontal ($h$) and kink ($k$) excitations are measured by
\begin{equation}
\begin{split}
    \Cth &= 2 \sum_{ij} \langle n_{i-1,j} n_{i,j} n_{i+1,j} \rangle, \\
    \Ctk &= 4 \sum_{ij} \langle  n_{i,j} n_{i+1,j} n_{i,j+1} (\id-n)_{i+1,j+1} \rangle.
\end{split}
\end{equation}
To identify the four-spin square, we measure:
\begin{equation}
    \Cf =  \sum_{ij} \langle  n_{i,j} n_{i+1,j} n_{i,j+1} n_{i+1,j+1} \rangle.
\end{equation}
The integer prefactors of $\Cth$ and $\Ctk$ account for the degeneracy of these excitations; the signal of $\Cth$ matches with the one of vertically oriented three-spin excitations, and the same for the four differently oriented three-spin kinks.
The $(\id - n)$ factor is necessary to disentangle the signals from larger particles. Without it, $\Ctk$ would also partially capture the four-spin particles described by $\Cf$. 

At finite $g$ these correlators do not measure the quasiparticles exactly, but only their approximations.
One way to amend this would be to adiabatically turn off the transverse field after the scattering~\cite{Bennewitz2025}, to reinstate the product-state nature of the excitations. 
In contrast with 1D cases, this approach does not seem feasible in our case. The timescale for an accurate reverse quench is much larger than the decay time of heavier products, and more importantly, the time needed for the products to reach the edge of the system.
Hence, we disentangle the scattering signal from the background by subtracting the value of the correlators at the end of the preparation procedure, at $tJ=10$.
The raw data without the subtraction is shown in Sec.~\ref{sec:correlations_error}.

The time evolution of the correlators is shown in Fig.~\ref{fig:scattering}(c).
After an initial ramp up coming from the preparation, the peak signals the creation of intermediate states.
The $\Cth$ and $\Ctk$ increase simultaneously, while the growth of $\Cf$ comes slightly later, and at the expense of $\Ctk$.
This shows that the four-spin particles are not directly created in the initial impact (which would require a coherent process in the fourth order of $g$) but through a resonant transition from the three-spin kink excitations~\cite{Balducci2022}.

The $g$ dependence of the peak amplitudes of the correlators is shown in Fig.~\ref{fig:scattering}(d). 
The peak of $\Ctk$ follows $\Cth$ for small $g$, while for $g/J \gtrsim 1.5$ the heights of the $\Ctk$ and $\Cf$ peaks match. 
This separates the two scattering regimes. 
At small to intermediate $g$ the particle size drives the particle creation: the creation of four-spin particles requires a higher order process and is thus suppressed. 
At large $g/J$ the interplay between the kinks and squares stabilizes the four-spin excitation.
This is the heavy excitation in the center of the lattice at large $g/J$.
Further evidence for multiple processes can be found in the decay characteristics of the correlators. 
Whereas it is monotonous at $g/J=1.5$, a kink emerges in the tail of the decay curve at larger values of $g$.
This suggests the involvement of multiple species of excitations decaying at different rates.

Being only averaged pictures of the evolution, the energy density and sums of local correlations do not provide information about the individual scattering channels.
To disentangle these, we measure long-range correlations between distant points in the lattice. 

We measure two-point correlations on the scattering axis (A) and in the orthogonal direction (O). 
In the $24 \times 24$ lattice, they are:
\begin{equation}
\begin{split}
    c_A &= \langle n_{7,7} n_{18,18} \rangle - \langle n_{7,7} \rangle \langle n_{18,18} \rangle, \\
    c_O &= \langle n_{7,18} n_{18,7} \rangle - \langle n_{7,18} \rangle \langle n_{18,7} \rangle,
\end{split}
\end{equation}
and the correlation between them:
\begin{equation}
    c_{AO} = \langle n_{7,7} n_{7,18} \rangle - \langle n_{7,7} \rangle \langle n_{7,18} \rangle.
\end{equation}
Three-particle processes are captured by a three-point correlator with one point in the center of the lattice, and two on the scattering axis or orthogonally: 
\begin{equation}
\begin{split}
    c_A^{(3)} &= \langle n_{7,7} n_{12,12} n_{18,18} \rangle - \langle n_{7,7} \rangle \langle n_{12,12} \rangle \langle n_{18,18} \rangle, \\
    c_O^{(3)} &= \langle n_{18,7} n_{12,12} n_{7,18} \rangle - \langle n_{18,7} \rangle \langle n_{12,12} \rangle \langle n_{7,18} \rangle.
\end{split}
\end{equation}

The evolution of the correlations is shown in Fig.~\ref{fig:scattering}(e), and $g$-dependence of peak amplitudes in Fig.~\ref{fig:scattering}(f).
At $g/J=1$, the single dominant peak is a signature of the elastic process.
It splits into two with increasing $g$, and at $g/J \gtrsim 1.5$ we clearly find two separate contributions. 
The first still corresponds to the elastic process, while the second signals the delayed arrival of particles generated through an intermediate resonance.
The second peak is accompanied by a peak in $c_O$.
The peaks of $c_A$ and $c_O$ are not equal because the magnon dispersion is not exactly sinusoidal, and thus the scattering is not precisely at the van Hove singularity. Note that the products are more concentrated in the off-axis direction for $g/J=1.5$.
We expect that this is accompanied by a finite-size effect, as the products are not point-like particles, but rather extended objects on the lattice. 
The negative $c_{AO}$ that accompanies the overlap of $c_A$ and $c_O$ shows that the presence of particles on the two axes is anti-correlated. This confirms that these are two separate two-particle processes, and not a four-particle process. 

As $g$ is increased further the second peak in $c_A$ -- and with it the $c_O$ contribution -- decreases, while the amplitude of the dominant peak increases.  
Simultaneously we observe an increase in the three-particle correlators, a signature of a three-particle process.
These by design measure a process where two particles propagate outwards with equal momenta, and the third remains static at the collision point.
We also measured the four-point correlator, but it does not deviate from zero in the presented range of $g/J$ (not shown).

In conclusion, we presented evidence of three qualitatively different scattering regimes. 
Elastic scattering dominates at small $g/J \lesssim 1.0$. 
At intermediate $g/J \sim 1.5$ it is joined by a process where an intermediate heavier particle --- a mixture of $\Cth$ and $\Ctk$ --- is created, and symmetrically decays into pairs of magnons.
At still larger $g/J \sim 2.0$ we find that the elastic contribution disappears, and is replaced by a three-particle process; the creation of a an excitation composed of $\Ctk$ and $\Cf$ along with two outwards propagating magnons. 

\section{Scattering in a false vacuum}
\label{sec:false_vacuum}

\begin{figure*}
    \centering
    \includegraphics[width=\textwidth]{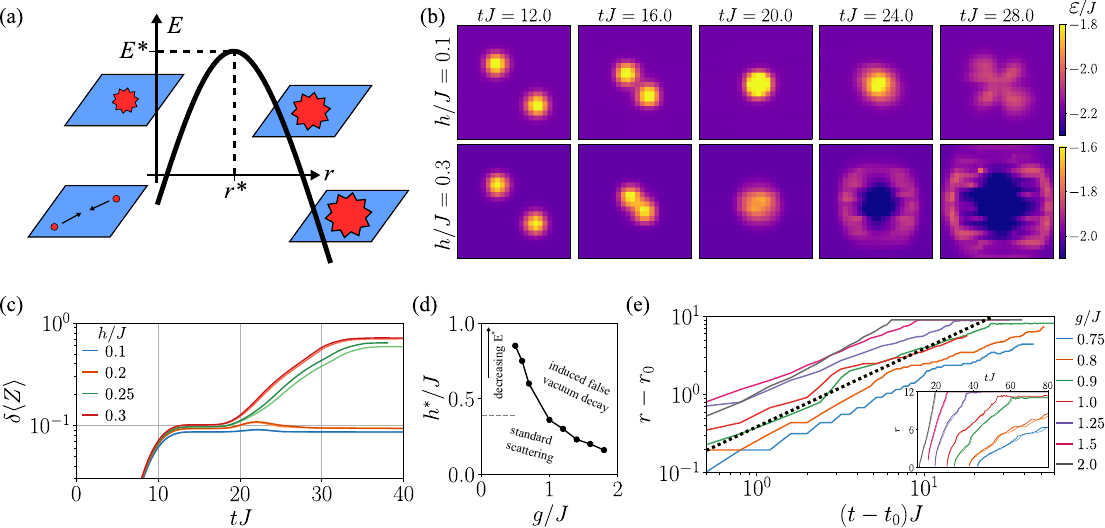}
    \caption{
    Scattering in the false vacuum.
    (a) A sketch of the energy landscape of a true vacuum bubble with radius $r$.
    (b) The evolution of the energy density during scattering with $h/J = 0.1$ and $h/J = 0.3$, both with $g/J=1.5$ and $\veck = \pipi$. 
    (c) The change in magnetization per site with varying $h$ for $g/J=1.5$ and $\veck = \pipi$.
    (d) The dependence of the threshold value $h^*$ on $g/J$. The gray dashed line corresponds to the threshold $h^*$ expected for $g\rightarrow 0$. See text for details.
    (e) The evolution of the radius of the true vacuum bubble in log-log scale for varying $g$, with some longitudinal field larger than the threshold value. We subtract the time $t_0$ when the radius reaches $r_0 = 3$ on the horizontal, and $r_0$ on the vertical axis. 
    The black dashed line corresponds to linear growth and acts as a guide to the eye.
    (Inset) The evolution of the radius in lin-lin scale. 
    In (c) and (e), the curves with different shades of the same color correspond to simulations with different bond dimension, darker is bigger. Here we use $\chi = 150$ and 200.
    }
    \label{fig:false_vacuum}
\end{figure*}

Next, we break the spin-inversion symmetry of the model with the longitudinal field, $h \neq 0$. This separates the spectrum into a stable and a metastable manifold, allowing us to study the stability of the metastable state to the scattering of magnons of the stable species.   
This is a setup amenable for studying false vacuum decay; how a metastable excitation-free state, \textit{a false vacuum}, decays into a truly stable configuration. 

Originating in cosmology and high-energy physics~\cite{Coleman1977}, the question has recently attracted attention in the context of metastability in quantum many-body systems.
It has been studied in 1D numerically~\cite{Lagnese2021, Sinha2021, Milsted2022, Lagnese2024, Abel:2020qzm, Abel2025, Abel:2025zxb} and experimentally~\cite{Zenesini2024, Zhu2024, Darbha2024, Vodeb_2025}. 
The microscopic mechanism behind false vacuum decay is probabilistic bubble nucleation~\cite{Lagnese2024}: the vacuum fluctuations induced through a quench are small bubbles of the true vacuum, where the system locally tunnels to the stable state.  
Assuming a spherical bubble with radius $r$, its energy consists of two contributions: a positive interface tension, and a negative bulk term coming from the energy difference between the two vacua~\cite{Coleman1977, Lagnese2024}. 
In 2D one expects the energy $E(r) = A \cdot 2J \cdot r - B \cdot 2h \cdot r^2$, with model-dependent geometrical factors $A$ and $B$. 
The parabola acts as a potential barrier between the false vacuum at $r=0$ and the true vacuum at $r\rightarrow\infty$, with the maximal energy at the critical bubble size $r^* \propto J/h$ and $E(r^*)\propto J^2/h$.
If a bubble bigger than $r^*$ is created, it can expand by transforming the potential energy of the false vacuum into the kinetic energy of the bubble's walls. The system thus transitions towards the stable manifold. 
The energy manifold is sketched in Fig.~\ref{fig:false_vacuum}(a).

Here, we prepare magnon wave packets in a false vacuum background, and study how their collision induces the transition into the stable manifold. 
In this case, we do not expect the system to fully tunnel \textit{through} the potential barrier as in unperturbed false vacuum decay, but rather \textit{over it}. 
Intuitively, the collision of the wave packets concentrates energy in the scattering site, and if the energy density is large enough, the system locally traverses the potential barrier and creates an expanding critical bubble.

Because the height of the potential barrier $E(r^*) \propto 1/h$, we can vary it by varying $h$ at fixed $g$. 
By doing this, we find two sharply separated dynamical regimes; see the evolution of energy density in Fig.~\ref{fig:false_vacuum}(b) for two representative values of $h$.
At the smaller $h$ we recover the process akin to the one in Fig.~\ref{fig:scattering}.
However, if $h$ is tuned beyond a threshold value, denoted $h^*$, the dynamics drastically change. 
Here, the scattering process generates a bubble of the true vacuum, which violently expands and encompasses the whole system.
This is reflected in the evolution of the magnetization, shown in Fig.~\ref{fig:false_vacuum}(c), exhibiting two types of behavior depending on $h$: it is approximately constant at $h < h^*$, and rapidly increases after scattering when $h > h^*$. 

In Fig.~\ref{fig:false_vacuum}(d), we show the dependence of the threshold $h^*$ on $g$. We obtain the points by simulating the scattering process with a set $g$ and varying $h$, and denoting $h^*$ as the smallest value of $h$ where we observe a large change of average magnetization. The solid line is a guide to the eye.
In the limit of $g\rightarrow 0$, where the model is classical, we can exactly determine the energies involved. 
By equating the energy of the wave packets $2 \times 8J$ with $E(r=r^*, h^*)$ (assuming a spherical bubble in 2D; $A = 2\pi$, $B=\pi$), we find $h^*(g\rightarrow0) \sim 0.39$, denoted by the gray dashed line in Fig.~\ref{fig:false_vacuum}(d).
Interestingly, this value strongly disagrees with the simulations.
This indicates that the creation of the bubble is not conditioned only on the energetics of the incoming particles, but also on the nature of their interaction, which depends on $g/J$.
In particular, the critical bubble is a many-body excitation, and $g/J$ needs to be large enough to enable its non-perturbative generation from the initial two-particle state.

Induced false vacuum decay has been studied in the context of quantum field theory in Refs.~\cite{Kuznetsov1997, Levkov2005, Demidov2015}.
These works study the process in the semi-classical limit, where it is shown that a collision of two particles does not trigger the decay: the scattering cross-section for such process remains exponentially suppressed up to infinite energy. 
A simple explanation is that while the scattering might carry enough energy to create the true vacuum bubble, it is a state of many excitations. Generating it from a two-particle state is an unlikely high-order process. 
This is in accordance with our results, which show that the process is suppressed at small $g/J$ even when enough energy is available.
We however find that increasing $g/J$ beyond the perturbative regime allows us to induce the critical bubble.

On the other hand, the concentration of the energy in the collision point locally heats the system, and the decay process might be related to false vacuum decay at finite-temperature~\cite{Linde1983}. 
This is a classical process in which thermal fluctuations bring the system over the potential barrier.  
However, this would mean only a weak dependence on $g/J$ (only due to the renormalization of the domain walls), whilst we observe a strong effect in Fig.~\ref{fig:false_vacuum}(d).

Additionally, the presented results are on a finite lattice, and frankly quite far from the continuum limit of field theory.
While scaling these simulations to much larger system sizes is a formidable task, we rule out the lattice effects by checking that the threshold $h^*$ does not depend on the size of the wave packet. 
See Sec.~\ref{sec:finite_size_wp}.

Finally, in Fig.~\ref{fig:false_vacuum}(e), we plot the evolution of the bubble radius for various $g/J$, with $h > h^*(g)$. 
The main panel shows a log-log plot where we subtract a $g$-dependent $t_0$ in the time axis, and $r_0=3$ in the vertical axis. The inset shows the raw data in linear scale.
The black dashed line corresponds to linear increase of $r$. It roughly matches the numerical data, implying that the bubble spreads ballistically, in agreement to a solitonic solution expected in QFT.
However, a precise quantitative statement is admittedly difficult to make.
The data is reasonably well converged in bond dimension (see inset, where we plot data obtained with different bond dimensions). Yet, the bubble, a coherent collective excitation, generates long-range entanglement, making it hard to accurately capture with tensor network methods. This can be seen from the appearance of the horizontal stripes in Fig.~\ref{fig:false_vacuum}(b) at later times, which are numerical artifacts.
Additionally, because of the finite lattice spacing, approximating the bubble as spherical might not be very accurate at initial times. 
For these reasons, we view the growth of the coherent, highly-entangled bubble as a promising target for experimental quantum simulation beyond the reach of classical techniques. 

In conclusion, we believe that the observed phenomenon is due to a combination of a local thermal and a quantum tunneling process. Further work is certainly required to completely understand their interplay.
It should be noted that it is not known whether Coleman's arguments and the critical droplet theory even apply in two dimensions~\cite{Moss_2025}. We plan to investigate this in an upcoming work.

\section{Discussion}

We studied the scattering of spin excitations in the 2D quantum Ising model using tree tensor networks.
The ordered phase of the Ising model is particularly well-suited for probing inelastic processes, because its low-energy excitations can be interpreted as composite bound states.
We characterized three distinct dynamical regimes: a perturbative elastic regime at small transverse fields; an intermediate regime where the dominant process produces pairs of magnons through an intermediate resonance; and a strongly inelastic regime at large transverse field, where we find a three-particle process involving the creation of two outward propagating particles along with an additional heavier excitation.
We used local correlations to identify bound states of different shapes and sizes, and disentangled different scattering channels by measuring long-range correlations. 

Next, we simulated the scattering of true vacuum bubbles in a metastable false vacuum.
We discovered a sharp dynamical transition between a conventional scattering regime and one where the collision induces a self-sustaining expansion of the true vacuum bubble.
We find that the threshold for inducing the false vacuum decay at small $g/J$ considerably exceeds the energy required to cross the potential barrier classically.
This leads us to believe that the generation of the critical bubble is a collective non-perturbative process.  
The long-time dynamics of the bubble expansion is hard to capture classically, making this an interesting setting for quantum simulation beyond the reach of classical methods. 
Probing the expansion of the bubble would have implications for understanding the false vacuum decay beyond the perturbative regime but, more broadly, is a probe of collective coherent phenomena in correlated systems and thermalization dynamics of quantum metastable states~\cite{Yin2025}.

From the methodological perspective, the presented work demonstrates that tree tensor networks can simulate scattering dynamics on sizeable 2D lattices.
This success is related to the fact that the entanglement generated in the scattering process is reasonably structured in a way that can be captured by a TTN.
This should hold true in a typical scattering setup, as long as the underlying vacuum is reasonably weakly entangled.
This leads us to believe that equivalent simulations could be performed in more complex models, opening the door for general simulations of scattering in two spatial dimensions.

Understanding scattering interactions in strongly-coupled lattice-gauge theories would elucidate mysteries related to the creation of matter and its confinement into composite states, while in condensed matter, scattering acts as a spectroscopic probe of the system and its excitations. 
There are interesting examples in the 2D Ising model already. The intertwined bands of vertical and horizontal 2-spin excitations can exhibit topological properties, which could be studied through the dynamics of wave packets~\cite{salerno2020interaction}.
In the Bose-Hubbard model, phonon scattering in the superfluid phase could be used to probe the Higgs amplitude mode, giving direct access to the Higgs-phonon coupling~\cite{endres2012}.

Finally, we see numerical experiments of scattering as a refined way to study quantum dynamics.  
This is typically done by introducing energy into the system through a global quench of the model's parameters.  
In this context, scattering is a way to inject the energy in a controlled and localized fashion, making it easier to track its evolution. This is particularly useful for studying inhomogeneous processes, such as the formation and decay of bound states or composite particles --- phenomena obfuscated by a global quench.

\section*{Acknowledgements}

LP thanks Lorenzo Maffi and Pietro Silvi for valuable discussions, Marco Rigobello and Roland Farrell for insightful comments on the manuscript, and Daniel Jaschke for support with software development. 
The tensor network simulations were performed with the \texttt{Quantum Tea} library~\cite{qtealeaves, qredtea, qredtea_benchmark}.
The exact diagonalisation calculations were performed with \texttt{QuSpin}~\cite{quspin}.

The work was supported by the European Union via ICSC - Italian Research Center on HPC, Big Data and Quantum Computing (NextGenerationEU Project No. CN00000013), EuRyQa (Horizon 2020), Horizon Europe program HORIZON-CL4-2022-QUANTUM-02-SGA via the project 101113690 (PASQuanS2.1), by the Italian Ministry of University and Research (MUR) via Quantum Frontiers (the Departments of Excellence 2023-2027); the World Class Research Infrastructure - Quantum Computing and Simulation Center (QCSC) of Padova University, and by the Istituto Nazionale di Fisica Nucleare (INFN): iniziativa specifica IS-QUANTUM.

The authors acknowledge computational resources of the INFN Padova HPC cluster, funded by the NextGenerationEU Terabit project on PNRR – Avviso n. 3264 "per il Rafforzamento e creazione di Infrastrutture di Ricerca", Missione 4, "Istruzione e Ricerca".

\section*{Data Availability}
The data plotted in the figures of this study are available in a Zenodo repository at \texttt{https://doi.org/10.5281/zenodo.18352073}.

\section*{Code Availability}

All numerical simulations in this study were performed using the \texttt{Quantum Tea} library. All features required to reproduce the figures and results presented in this work are available in the public release of the library.
The script used to generate the initial states is available in a Zenodo repository at \texttt{https://doi.org/10.5281/zenodo.18352073}.

\bibliography{bibliography}

\section*{Methods}
\label{sec:methods}
\subsection{Tree tensor networks}

The reason why the real-time simulations of scattering are limited to one spatial dimension is largely of methodological nature.
The most successful method for computing the time evolution of quantum states are tensor networks, namely the matrix product states (MPS). 
Due to its shape, the MPS is well adapted to describing one dimensional systems with nearest neighbor interactions, where the correlations are mostly short-ranged. 
More precisely, it is understood that MPS can efficiently encode states obeying the entanglement area law~\cite{Schollwock2011}.

To describe 2D systems with an MPS, the lattice is mapped back to one dimension using a space-covering curve. This induces effective long-range interactions, and consequently long range entanglement. Importantly, the range of these interactions scales with the system size. 
MPS is thus often not powerful enough to accurately describe states on large 2D lattices.

Tree tensor networks (TTN), which we use in this work, are an extension of the MPS. 
They are built out of rank-three tensors arranged into a binary tree. The leaves of the tree have two physical indices each, while the tensors in the upper layers act as auxiliaries used to capture long-range correlations and entanglement. 
The TTN benefits from a larger connectivity compared to the MPS. This reduces the maximal distance in the network between two physical sites from linear to logarithmic in the system size. 
All important algorithms that were developed for manipulating MPS can be generalized to TTN with little change in the computational complexity.

The TTN is still a one-dimensional network, just as the MPS. As such, the 2D system still has to be mapped to 1D, and the representability of the network is limited by the same 1D area law. The auxiliary tensors in the TTN play an important role here, as their presence means that the long-range interactions that emerge from the mapping can be captured more accurately.
Exploring the frontier of problems that can be simulated with TTNs is important for understanding which problems in quantum dynamics are completely out of reach, and which could be tackled with large-scale classical simulations.
The presented work is an example of such exploration. 
We find that TTNs are sufficient to accurately describe the post-scattering states up to large values of the transverse field, and for surprisingly large two-dimensional lattices.

\subsection{Numerical details}

We use tree tensor networks to encode the state of the system~\cite{Tagliacozzo2009, Murg2010}, and the time-dependent variational principle (TDVP)~\cite{Haegeman2016, Bauernfeind2020} to evolve them in time. 
We use the single-site version of TDVP, where each tensor is updated separately. This version of the algorithm cannot grow the bond dimension during the evolution, so our initial states start at the maximal bond dimension. This introduces some overhead in the initial steps, but we find that this approach performs best for evolving the system to very large times.
For simplicity we use the same time step for all simulations: $dt = 0.02/J$.

Each simulation is performed using one NVIDIA H100 GPU with 80 GB of RAM. 
Including the measurements, which we preform every ten time steps, one time step for the whole $24 \times 24$ system takes $\sim 0.7~\mathrm{min}$ at $\chi=100$, $\sim 2~\mathrm{min}$ at $\chi=150$, $\sim 3~\mathrm{min}$ at $\chi=200$.

As the tensors in the TTN are of rank three, they contain $\chi^3$ parameters. Comparing only the number of parameters, a TTN with $\chi=250$ (the biggest we use) corresponds to an MPS with $\chi \sim 2800$.  

\subsection{Preparation of wave packets}
\label{sec:preparation}

In this section, we describe how we prepare the initial states.
As shown in Fig.~\ref{fig:excitations_and_spectrum}, the manifold of single-spin excitations is reasonably well separated from the rest of the spectrum. It is also adiabatically connected to the product-state excitations at $g=0$.
We thus prepare wave packets at this non-interacting point, and adiabatically turn on $g$ towards the target value. 
As there are no level crossings, we expect the resulting states to be an accurate representation of the dressed wave packets at finite $g$.
This procedure could be performed experimentally on currently available quantum simulators. The central requirement is site-resolved control of the transverse field~\cite{Bennewitz2025}.

More precisely, an exact wave packet for $g=0$, centered around position $\vecr_0$ and momentum $\veck$ is 
\begin{equation}
    \ket{\vecr_0, \veck} = \frac{1}{N} \sum_{\vecr} f(\vecr, \vecr_0) e^{i \veck \cdot \vecr} X_{\vecr}\ket{0},
    \label{eq:product_wavepacket}
\end{equation}
where $\ket{0}$ is the $Z$-polarized product state, and the sum runs across all sites of the $N \times N$ lattice.
The envelope function $f$ is the Gaussian
\begin{equation}
    f(\vecr, \vecr_0) = \frac{1}{\sigma\sqrt{\pi}} e^{\frac{-(\vecr-\vecr_0)^2}{2\sigma^2}}.
\end{equation}

In practice, Eq.~\eqref{eq:product_wavepacket} represents a superposition of a few hundred product states, to be encoded in a TTN. 
As the bond dimension doubles for each sum of two tensor networks, performing this summation directly leads to an exponential explosion in required memory, and is thus not feasible.
It is possible to compress the state between the summations steps, but the result of such procedure would depend on the order of the summation and compression steps.

To circumvent this issue, we employ a variational algorithm to fit an initial tensor network state $\psi$ onto a sum of the form $\sum_i \alpha_i \psi_i$.
The algorithm, based on \textit{effective projectors}, a generalisation of effective operators, is described in the next section.
The procedure is computationally cheap and allows us to efficiently sum hundreds of tensor network states, especially if they are simple product states like the summands in Eq.~\eqref{eq:product_wavepacket}. 
We find numerical errors (mismatch between $\alpha_i$ and $\langle \psi \vert \psi_i \rangle$) on the order of $\sim 10^{-14}$, and thus completely negligible~\footnote{During the preparation of the manuscript, it was brought to our attention that a single excitation state like Eq.~\eqref{eq:product_wavepacket} can be exactly represented as an MPS with bond dimension two. The most efficient way to generate the initial product state of two wave packets is thus to write the operator that generates each wave packet as an MPO (each with $\chi=2$), and contract them with a TTN representation of $\vert 0 \rangle$ ($\chi=1$). }.

Once we obtain the desired state at $g=0$, we adiabatically increase the transverse field following 
\begin{equation}
    g(t) = g^* \sin^2  \left(\frac{\pi}{2} \sin^2 \left(\frac{\pi}{2} \frac{t}{\tau} \right)\right)   
\end{equation}
with $g^*$ the target value, and $\tau$ the quench timescale. 
This pulse shape is commonly used for adiabatic quenches due to its nice analytical properties~\cite{Albash2018}.
In particular, it is important that the ramp is smooth (differentiable) in the whole interval, including its edges.
We found that a linear ramp $g(t) = g^* t / \tau$ induces a considerable density of spurious vacuum fluctuations at its non-differentiable points at $t=0$ and $t=\tau$.

\begin{figure}
    \centering
    \includegraphics[width=\columnwidth]{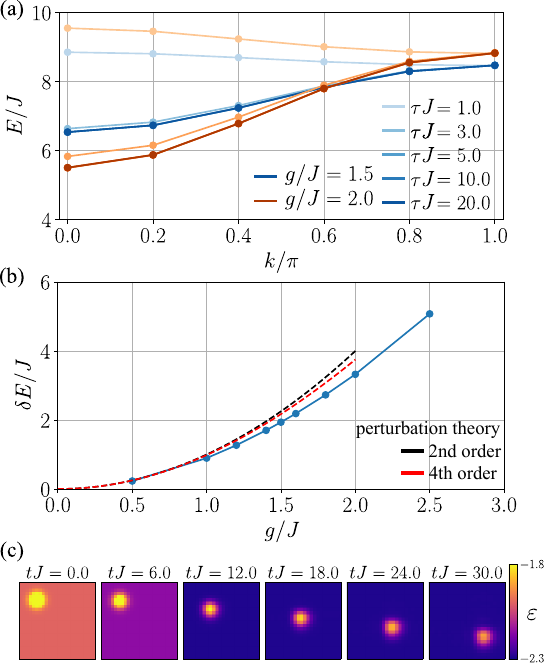}
    \caption{
    Preparation of a wave packet.
    (a) The dispersion obtained with the adiabatic quench to two values of $g$ (colors) with different $\tau$ (shades). The wave packet has width $\sigma=2$ and momentum $(k,k)$.
    (b) The dependence of the bandwidth, $\delta E = E(k=\pi)-E(k=0)$, on $g$. The dashed lines are perturbative results, black for second (Eq.~\eqref{eq:dispersion}) and red for up to fourth order.
    (c) The evolution of the energy density for a wave packet generated with $\tau J=10$, $\veck = (\frac{\pi}{2}, \frac{\pi}{2})$ and $g/J=1.5$. 
    }
    \label{fig:preparation_of_wavepackets}
\end{figure}

Fig.~\ref{fig:preparation_of_wavepackets}(a) shows the dispersion of a wave packet with momentum $\veck = (k, k)$. 
The excitation energy is obtained by performing the same quench on the polarized ground state, and subtracting the resulting energy. 
It is safe to assume that the quench is sufficiently close to the adiabatic limit when the dispersion converges with increasing $\tau$. As expected, larger $\tau$ is required closer to the critical point, as the gap becomes smaller. However, both curves are reasonably well converged for $\tau J \gtrsim 5$. 
Given these results, we stay on the safe side and fix $\tau = 10/J$ as the state preparation timescale.

As an interesting aside, note that a very short quench time (see $\tau J = 1$) results in a qualitatively different dispersion curve.
This implies that a sudden quench from a product state ($\tau=0$) might lead to qualitatively wrong results.

The width of the obtained band is shown in Fig.~\ref{fig:preparation_of_wavepackets}(b), where $\delta E = E(\veck=\pipi)-E(\veck=\mathbf{0})$. The dashed lines are predictions from perturbation theory, up to the second (black) and up to the fourth (red) order in $g/J$. 

We follow the propagation of a single wave packet through the evolution of the energy density in Fig.~\ref{fig:preparation_of_wavepackets}(c).
We find a well-behaved wave packet with minimal dispersion of its width.

To suppress lattice effects, the wave packets should be much bigger than the lattice constant. 
In practice, size of the wave packets is limited by the combination of the finite system size and the time of the adiabatic ramp. 
The wave packets start moving during the ramp, and it is critical to reach the target value of $g$ before they collide.
In the simulations presented in this work we set the size of the wave packets to  use $\sigma = 2$ in units of the lattice constant.
In practice this means that a wave packet noticeably changes the energy density in an area of approximately $8\times 8$ lattice sites. 
We find that the lattice effects qualitatively change the propagation of a single wave packet only as $\sigma \sim 1$, and the extent of the wave packet becomes comparable to a single unit cell.
In Sec.~\ref{sec:finite_size_wp} we test the effect of changing $\sigma$ on the induced false vacuum decay. 

Finally, we wish to point out the existence of alternative methods for preparing wave packets.
There is a considerable body of work discussing the search for excited states with tensor networks, including ones with well-defined momentum~\cite{Haegeman2012, Haegeman2013, Li2024, Osborne2025}.
A superposition of such states with different momenta can be inserted into a vacuum state to construct numerically exact wavepackets, and study scattering processes~\cite{Vanderstraeten2014, VanDamme2021_scattering, Milsted2022}.
However, these methods predominantly consider one dimensional spin chains encoded in MPS, or their cousin from the thermodynamic limit, the iMPS.
As such, they rely on the translational symmetry of the MPS ansatz.  
Generalizing them to two dimensional lattices is possible for a certain mapping of an infinite cylinder to MPS~\cite{VanDamme2021_2d}. 
However, the structure of the TTN breaks the translational invariance that is present in the MPS. Generalizing the excitation ansatz to generic mappings, 2D square systems on a torus, as well as to TTNs, is not straightforward. 
We leave this step for future work.

\subsection{Efficient summation of TTN states}
\label{app:effective projectors}

Here, we describe an algorithm used for the approximate summation of TTN states.
The algorithm is implemented in the \texttt{Quantum TEA} library as $\texttt{TTN.sum\_approximate()}$.

The goal is the following: given a set of tensor networks states $\phi_i$ and corresponding amplitudes $\alpha_i$, efficiently construct a tensor network representation of the state $\psi = \sum_i \alpha_i \phi_i$.

\begin{figure}[t]
    \centering
    \includegraphics[width=\columnwidth]{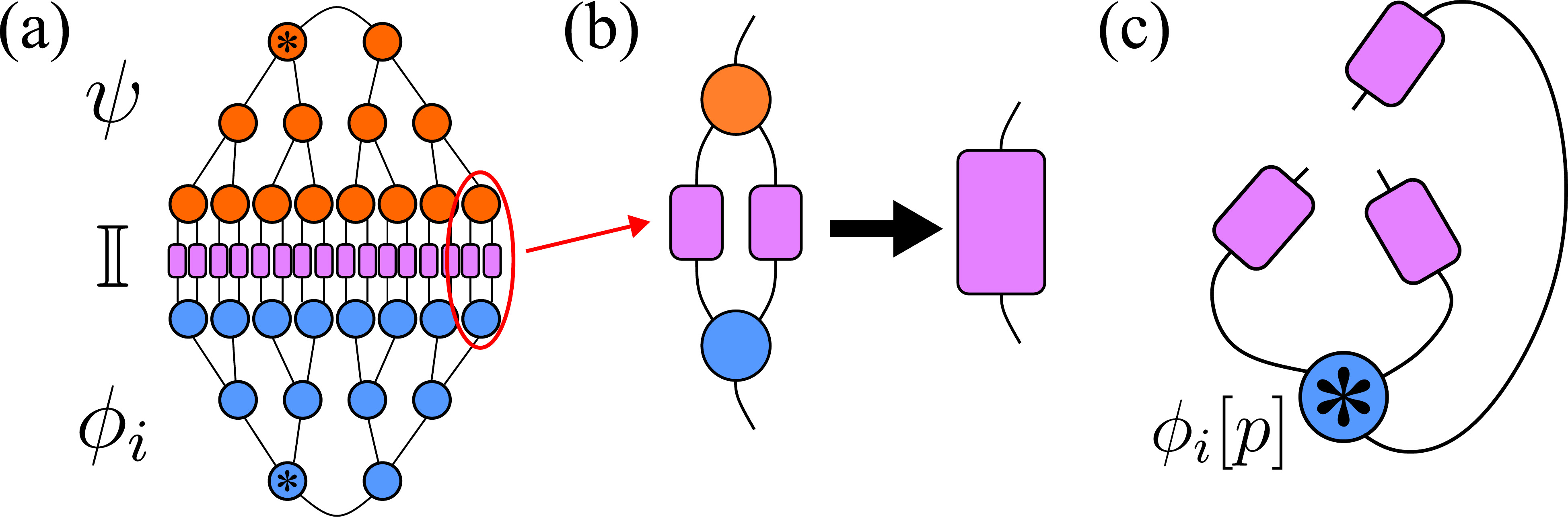}
    \caption{
    Summation of TTN states.
    (a) A schematic representation of a tensor network corresponding to an overlap between two TTN states; $\langle \psi \vert \mathds{1} \vert \phi_i \rangle$.
    (b) One step of the iterative calculation of the effective operators. The two effective operators and the tensors from $\psi$ and $\phi_i$ are contracted into a new effective operator, which lives one layer higher. 
    (c) The operator representing the effective projection of $\phi_i$ onto $\psi$.
    }
    \label{fig:effective_projectors}
\end{figure}

We start by choosing a random TTN $\psi$, and constructing a tensor network representing $\langle \phi_i \vert \psi \rangle$ for each $\phi_i$, see Fig.~\ref{fig:effective_projectors}(a). 
The orange TTN represents $\psi$, and the blue corresponds to $\phi_i^*$. The star symbol denotes the isometry centers.
The pink rectangles in between are an identity operator in the MPO form. Its presence is not necessary, but we include it for clarity. 
Next, we build the effective operators on the links of the upper TTN. This is done by iteratively contracting the pink operators and the corresponding tensors of $\psi$ and $\phi_i$, as shown in Fig.~\ref{fig:effective_projectors}(b), starting in the physical (lowest) layer of the network and moving towards the isometry center of the upper tree $p$. 
At the end of the iterative contraction, we have the effective operators surrounding the isometry center $p$ in $\psi$, together with the corresponding tensor $\phi_i [p]$, shown in Fig.~\ref{fig:effective_projectors}(c).
Contracting this object with $\psi[p]$ would give $\langle \phi_i \vert \psi \rangle$.

Instead, we interpret this object as a local representation of $\langle \phi_i \vert^p$. The superscript $p$ denotes that the object refers to position $p$ in the network. 
As these are small three-legged tensors, they can be efficiently summed. 
We hence assign the linear combination of these objects for all $\phi_i$ as the tensor in position $p$ of $\psi$:
\begin{equation}
        \psi[p] \leftarrow \sum_i \alpha_i \vert \phi_i \rangle^p. 
\end{equation}
By sweeping through all positions $p$ as one would in a standard DMRG, updating the effective operators and assigning new tensors, we obtain an approximate representation of $\sum_i \alpha_i \phi_i$. 
For the same reasons as in DMRG (we are working with effective operators), the procedure does not necessarily converge after the first sweep. In this case, iterative repetitions of the sweeps are necessary.
This is typically the case if the summand states $\phi_i$ contain long-range features.

In practice, the building blocks of our wave packets predominantly contain local features, and we typically achieve very good accuracy with a single sweep. 
Assuming that $\phi_i$ are orthogonal, a good measure of error is $\epsilon = \sum_i \vert \alpha_i - \langle \psi \vert \phi_i\rangle\vert$. In our simulations, we find $\epsilon \sim 10^{-14}$ even for initial $\psi$ with a small bond dimension of $\chi = 30$. 

\subsection{Errors induced by the state preparation}
\label{sec:correlations_error}

The sketch of product-state excitations in Fig.~\ref{fig:excitations_and_spectrum}(a), as well as the definition of correlators $\Cth$, $\Ctk$ and $\Cf$ defined in the main text, is only exact at $g=0$, where the eigenstates of the system are product states. 
As $g/J$ is increased these states mix, which can be interpreted as the appearance of quantum fluctuations in the vacuum, and the dressing of the quasiparticles with higher excitations. 
However, as long as the system is in the ordered phase, the correlations between the $Z_i$ operators remain a good, albeit approximate measure for the presence of excitations.

\begin{figure}[t]
    \centering
    \includegraphics[width=\columnwidth]{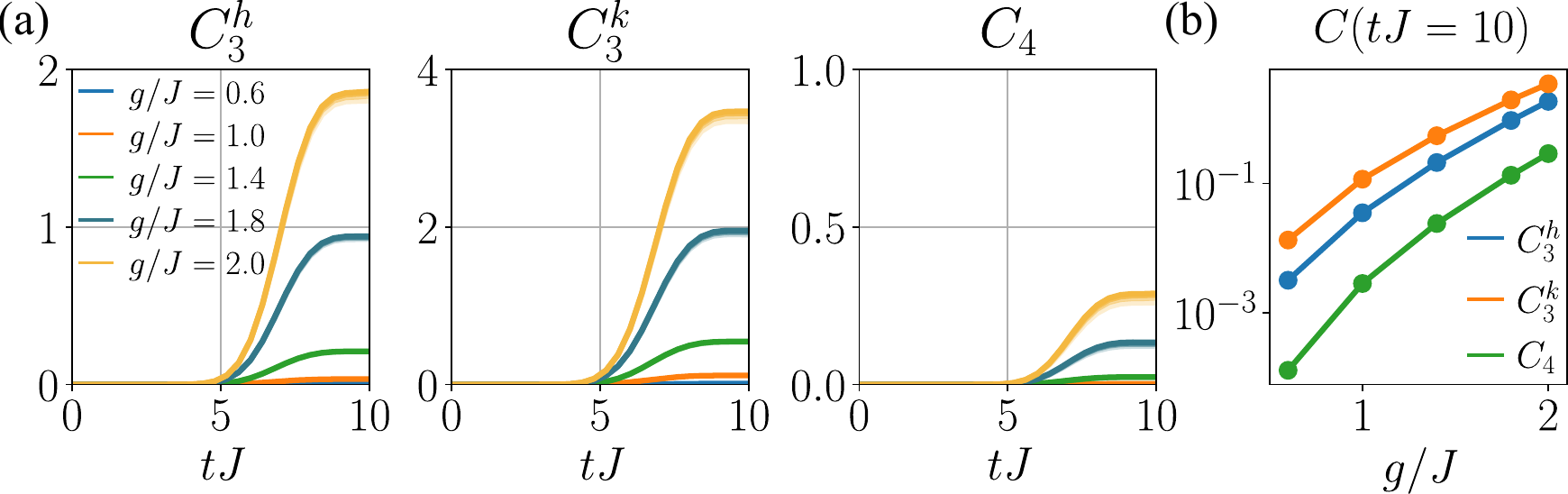}
    \caption{
    Measure of the error induced by the adiabatic preparation.
    (a) The evolution of the raw values of the correlations $\Cth$, $\Ctk$ and $\Cf$ during the preparation with $\tau J = 10$. 
    (b) The value of the correlations at the end of the state preparation versus $g/J$. 
    }
    \label{fig:correlation_error}
\end{figure}

In the main text, we isolate the effects of the scattering by subtracting the values of each correlator at the end of the preparation. 
In Fig.~\ref{fig:correlation_error} we show how the correlators increase during the state preparation procedure, and how this depends on the value of $g/J$.
The state is a two-wavepacket state, studied in Fig.~\ref{fig:scattering}.
In Fig.~\ref{fig:correlation_error}(a) we plot the increase of $\Cth$, $\Ctk$ and $\Cf$, while in Fig.~\ref{fig:correlation_error}(b) we show the dependence of the value at the end of the quench ($tJ=\tau$) on $g/J$.

The increase in the correlators originates from two effects; the vacuum fluctuations and the dressing of the wave packets. 
The former effect is extensive in the system size, while the latter scales with the number of the wave packets in the system.

\subsection{Finite size effects: the effect of wave packet size on the induced decay of the false vacuum}
\label{sec:finite_size_wp}

If one wishes to apply a quantum-field-theoretic interpretation to the generation and explosion of a true vacuum bubble shown in Fig.~\ref{fig:false_vacuum}, it is necessary to verify that the observed phenomenon is not a finite-size or lattice artifact. 
In practice, we do not think it is feasible to significantly increase the system size much beyond what is considered in this work. 
We therefore at this point cannot address the effects coming from the finite size of the system itself. 
However, lattice effects can be ruled out by checking that the phenomenon is independent of the wave-packet size.
If that were not the case, we would have expected a strong dependence of both the evolution of the magnetization as well as the threshold value $h^*$ on $\sigma$.

\begin{figure}[t]
    \centering
    \includegraphics[width=\columnwidth]{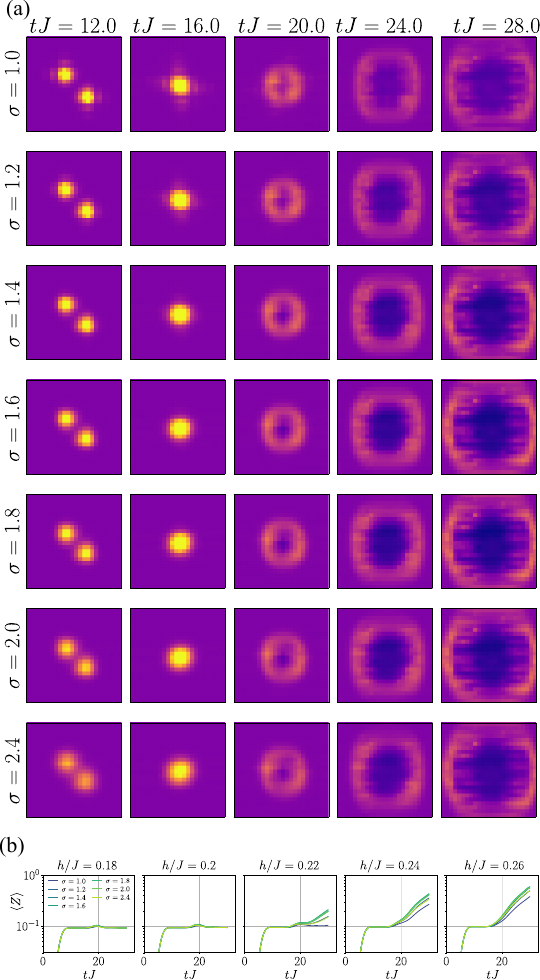}
    \caption{
    Dependence of the scattering in the false vacuum on the size of the wave packets.
    (a) The evolution of the energy density through the scattering with varying wave packet size $\sigma$. We set $g/J = 1.5$ and $h/J =  0.26$, above the decay threshold.
    (b) The evolution of the magnetization for varying $\sigma$ and $h/J$. We keep $g/J=1.5$.
    }
    \label{fig:wp_size}
\end{figure}

In Fig.~\ref{fig:wp_size}, we present the simulations of scattering at $g/J = 1.5$ for a range of wave packet sizes $\sigma$. We focus on the regime close to the threshold value of $h^* \sim 0.25$. 
In Fig.~\ref{fig:wp_size}(a) we show the evolution of the energy density during the scattering process at $h/J=0.26$, above the threshold. 
In the range of sizes we show here, the observed phenomenon does not change qualitatively. 
Strong lattice effects are only visible in the incoming wave packets at the smallest values of $\sigma =1.0$ and $1.2$.

In Fig.~\ref{fig:wp_size}(b) we show the evolution of the magnetization for a set of $h/J$ and a range of $\sigma$.
Again, we find that the dependence on $\sigma$ is negligible, with the exception of the smallest values.
These plots also confirm that the observed threshold $h^*$ does not drift with changing $\sigma$.

\end{document}